\documentclass{raa}            

\usepackage{graphicx,times}             

\begin{document}

   \title{The spin-up of a star gaining mass in a close binary system on the thermal time scale
}

   \volnopage{Vol.0 (20xx) No.0, 000--000}      
   \setcounter{page}{1}          

   \author{Evgeny Staritsin
   }

   \institute{K.A. Barkhatova Kourovka Astronomical Observatory, B.N. Yeltsin Ural Federal University,
             pr. Lenina 51, Ekaterinburg 620000, Russia; {\it Evgeny.Staritsin@urfu.ru}\\
   }

   \date{Received~~2009 month day; accepted~~2009~~month day}

\abstract{ We investigate the exchange of mass in a binary system as a channel through which a Be star can receive a rapid rotation.
The mass-transfer phase in a massive close binary system in the Hertzsprung-gap is accompanied by the spinning up of the accreting component.
We consider a case when the mass of the accreting component increases by 1.5 times. The component acquires mass and angular momentum while in a state of critical rotation.
The angular momentum of the component increases by 50 times. Meridional circulation effectively transports angular momentum inside the component during the mass-transfer phase
and during the thermal time scale after the end of the mass-transfer phase. As a result of mass transfer, the component acquires the rotation typical of classical Be stars.
}

   \authorrunning{E. Staritsin }            
   \titlerunning{The spin-up of a star gaining mass }  

   \maketitle

%
%
\section{Introduction}           
\label{sect:intro}

Classical Be type stars include B stars in whose spectra emission in the Balmer lines of hydrogen have been recorded (Porter \& Rivinius \cite{PR03}). These stars are not supergiants
and demonstrate fast rotation. The rotation velocities of Be stars in the early spectral subclass are in the range of 60\%-100\% of the Keplerian value (Cranmer \cite{C2005}). The origin
of the rapid rotation of Be stars is still unclear. A star can begin to rapidly rotate as a result of some different processes. Currently, mass transfer in a close binary system is
regarded as the most likely reason for the creation of Be type stars (Pols et al. \cite{PCWH91}; Portegies Zwart \cite{PZ95}; Van Bever \& Vanbeveren \cite{BV97}; Shao \& Li \cite{SL14};
Hastings et al. \cite{HLWS21}).

The evolution of the more massive star in a close binary system is faster than the evolution of the less massive star. The radius of the more massive star increases until its outer
layers begin to fall into the less massive star due to gravity. Falling matter swirls around the less massive star. Part of the falling matter and the angular momentum contained within
is attached to a less massive star. If the acquired angular momentum is instantly distributed over the star's entire volume, the star receives the rotation typical of Be stars after
an increase in its mass by 5\% -10\% (Packet \cite{P1981}).

The real transfer of angular momentum occurs over time. The transfer of angular momentum in the interiors of the accreting components of binary systems by meridional circulation
and shear turbulence is considered (Staritsin \cite{St19}, \cite{St21}) in the framework of Zahn (\cite{Zahn92}) model. In one case, a B star accretes mass while in the envelope
of a red supergiant (Staritsin \cite{St21}).
We also considered mass transfer in the Hertzsprung-gap of the HR diagram, when a more massive star fills its critical Roche
lobe after the end of evolution on the main sequence. We considered a special case where part of the mass-transfer phase lasts for the nuclear time scale of helium burning
of the more massive star (Staritsin \cite{St19}).
In both cases, the less massive star accretes some mass. By the end of accretion,
the surface rotation velocity increases to the Keplerian value. After the end of accretion, the transfer of angular momentum from the outer layers of the accretor to the inner
layers continues, and the surface rotation velocity becomes less than the Keplerian one. In both cases, the main mechanism of angular momentum transfer is meridional circulation.
In both cases, the star with accreted mass has the rotation typical of Be stars for some time, subject to a number of conditions.

This article discusses the possibility of creating a Be star in the general case of mass transfer in a close binary system in the Hertzsprung-gap on the thermal time scale.
Accretion into a star in a state of critical rotation is allowed. The angular momentum transfer in the depths of the accretor is carried out by meridional circulation and is considered
in the framework of the Zahn (\cite{Zahn92}) model.

\section{Basic equations and simplifications}
\label{sect:basic}
\subsection{Accretor mass increase}
\label{subsect:mass}
Mass transfer in binary systems has been studied for over half a century, beginning with Crawford (\cite{C1955}). At present, the mass lost by the initially more massive component during
the filling of the Roche lobe can be reliably determined. The mass and angular momentum gained by the other star of the pair, as well as the mass and angular momentum lost from
the system and the manner in which this occurs, remain the target of research.

There is a lot of evidence for nonconservative mass transfer in binaries with B spectral subclass components. For example, nonconservative mass transfer reproduces the observed
distribution of Algols over periods and is in better agreement with the mass ratio distribution of Algols (Van Rensbergen et al. \cite{RGM11}). To achieve agreement between
the calculated and observed distributions, a nonconservative mass-transfer phase is primarily required in binary systems with components of early spectral subclasses B.
The infrared excesses and strong Balmer continua observed in some interacting binary stars with the component of the early spectral subclass B also indicate nonconservative mass
transfer (Deschamps et al. \cite{DBJ15}). Despite this and other evidence of nonconservative mass transfer, a quantitative theory of mass and angular momentum loss from a close binary
system has not been developed. Therefore, a rough description of nonconservative mass transfer in population synthesis calculations is used. For example, the observed abundance
of Be stars of the early spectral subclass in the Galaxy can be reproduced under the assumption that half of the mass lost by the initially massive component is lost from
the system (Shao \& Li \cite{SL14}).

We considered mass transfer in a close binary system with the period of $P{=}35^d$  and component masses of $13.4\,M_\odot$ and $10.7\,M_\odot$ under this assumption
(Shao \& Li \cite{SL14}). By the time the initially massive component fills the Roche lobe, the secondary star is rotating synchronously with the orbital rotation.
The secondary star angular velocity is $(\Omega_s)_0{=}2\times10^{-6}\:\mbox{s}^{-1}$, the surface rotation velocity at the equator is $10\:\mbox{km}\,\mbox{s}^{-1}$.
The massive component is losing $10.5\,M_\odot$ at an average rate of $8.8\times10^{-4}\:M_\odot\,\mbox{year}^{-1}$. The secondary star accretes half the mass lost
by the massive component. We calculate accretion to the secondary star at a constant rate of $\sim4.4\times10^{-4}\:M_\odot\,\mbox{year}^{-1}$ over a period of $1.2\times10^4\:\mbox{years}$.
An increase in the secondary star's mass is accompanied by an increase in the mass of the convective core and an increase in its hydrogen content.
We assumed that semi-convection in the layer above the convective core is so effective that the distribution of hydrogen in this layer is the same as for a single star with
the same mass and hydrogen content in the convective core. Upon completion of the mass exchange, the component masses are $2.9\,M_\odot$ and $16\,M_\odot$, respectively.
Assuming that the matter lost from the system will carry away the secondary star's orbital angular momentum (Shao \& Li \cite{SL14}), we obtain an estimate of the period
of the system after the end of the mass exchange: $P\approx470^d$.

\subsection{Accretor angular momentum increment}
\label{subsect:increment}
The angular velocity of the accreted matter should coincide with the angular velocity of the star's surface. This condition is taken into account in problems of mass and angular
momentum accretion onto a star from a Keplerian disk (Paczynski \cite{P1991}; Popham \& Narayan \cite{PN91}; Colpi et al. \cite{CNC91}; Bisnovatyi-Kogan \cite{BK93}). If the star's
rotation velocity is less than the Keplerian one, the rotation velocity of the matter in the disk should decrease from the Keplerian value to the rotation velocity of the star's
surface in the narrow boundary layer near the star. The star-boundary layer-disk system was investigated by Paczynski (\cite{P1991}) and Popham \& Narayan (\cite{PN91}) with the $\alpha$-viscosity model (Shakura \& Sunyaev \cite{SS73}). In the obtained stationary solutions, the angular momentum entering the star is close to the Keplerian moment determined
near the surface of the star, at any velocity of the star's rotation.

In the case of a star's critical rotation, the angular rate of rotation in the disk decreases monotonically with distance from the star according to Kepler's law. A star in a state
of critical rotation can accrete mass due to the removal of angular momentum by shear stresses in the accretion disk (Paczynski \cite{P1991}; Popham \& Narayan \cite{PN91}; Colpi et al. \cite{CNC91}). Later, Bisnovatyi-Kogan (\cite{BK93}) showed that a star in a state of critical rotation can accrete both mass and angular momentum from the disk, while moving along a sequence of stars of increased mass with critical rotation.

Contemporary accretion disk research is aimed at searching for physical mechanisms for the transfer of mass, energy, and angular momentum. According to Belyaev \& Rafikov (\cite{BR12})
and Coleman et al. (\cite{CRP22}), waves can efficiently transfer the angular momentum from the narrow boundary layer to both the star and disk. No stationary solution is implemented
in this case, and the rate of angular momentum input to the star can greatly vary with time over several rotation periods of the inner edge of the disk (Belyaev et al. \cite{BRS12};
Belyaev et al. \cite{BRS13}; Hertfelder \& Kley \cite{HK15}).

We adopted the following recipe for determining the angular momentum input to the accretor. Let $\Omega_s$ be the angular velocity of the accretor surface, $\Omega_c$ be the angular
velocity of the inner edge of the Keplerian part of the disk. At the beginning of mass exchange, $\Omega_s<\Omega_c$. By analogy with the Keplerian accretion disk (Paczynski \cite{P1991}; Popham and Narayan \cite{PN91}), we will assume that the mass is added to the accretor with an angular velocity $\Omega_s$. The angular momentum at the accretor boundary is determined
by the angular velocity of rotation of the inner edge of the disk $\Omega_c$. Thus, the angular momentum enters the star along with the mass and in the form of an additional term due to turbulence, waves, or another mechanism not yet considered:
\begin{eqnarray}
\frac{dJ}{dt}&=&\frac{2}{3}R^2(\Omega_{c}-\Omega_{s})\dot M, 
\label{eq001}
\end{eqnarray}
where $J$ - angular momentum of the accretor, $R$ - accretor size, and $\dot M$ - mass accretion rate.

After the angular velocity of the accretor surface’s rotation becomes equal to the critical value, the angular momentum flux due to turbulence stops. Until the end of the mass exchange,
the evolution of the accretor is calculated on the assumption that the rate of rotation of its surface is equal to critical:
\begin{eqnarray}
\Omega_{s}=\Omega_{c}. 
\label{eq002}
\end{eqnarray}
According to the results of Bisnovatyi-Kogan (\cite{BK93}), the accretor should move along a sequence of stars of increased mass with a critical rotation.

\subsection{Angular momentum transfer in the accretor}
\label{subsect:transfer}
The angular momentum transfer in a star's radiative layers is taken into account in the framework of the shellular rotation model (Zahn \cite{Zahn92}). This model considers
two angular momentum transfer mechanisms, meridional circulation and shear turbulence. The transport properties of turbulence in the horizontal direction (that is, along
the surface of constant pressure) are much more pronounced than in the vertical direction. Therefore, an arbitrary surface of constant pressure rotates almost in a solid state.
The angular velocity of rotation can vary in the vertical direction. The transfer of angular momentum is described by the law of the conservation
of angular momentum (Tassoul \cite{T1978}):
\begin{eqnarray}
\label{eq003}
\frac{\partial(\rho\varpi^2\Omega)}{\partial t}+
\mbox{div}(\rho\varpi^2\Omega{\bf u})
=\mbox{div}(\rho\nu_{\mbox{v}}\varpi^2\mbox{grad}\Omega).  
\end{eqnarray}
The rate of meridional circulation $\bf u$ is determined from the law of the conservation of energy in a stationary form (Maeder \& Zahn \cite{MZ98}):
\begin{eqnarray}
\label{eq004}
\rho T{\bf u}\mbox{grad}s=\rho\varepsilon_n+
\mbox{div}(\chi\mbox{grad}T)-\mbox{div}{\bf F}_h.  
\end{eqnarray}
The equations are solved taking into account the first order of smallness in the expansion of the vertical component of the meridional circulation velocity in latitude:
$U_r(r,\theta)=U(r)\mbox{P}_2(\theta)$ (Zahn \cite{Zahn92}).
Here $U(r)$ - the amplitude of the vertical component of the meridional circulation velocity (hereinafter, the amplitude of the meridional circulation velocity).
In these equations
$\mbox{P}_2(\theta)$ - the associated Legendre function of the second kind,
$\rho$ - density,
$\varpi$ - distance to the axis of rotation,
$\nu_{\mbox{v}}$ - turbulent viscosity in the vertical direction,
$T$ - temperature,
$s$ - specific entropy,
$\varepsilon_n$ - nuclear energy release rate,
$\chi$ - thermal conductivity,
${\bf F}_h$ - turbulent enthalpy flow in the horizontal direction:
${\bf F}_h=-\nu_h\rho T\partial{s}/\partial{\bf i_\theta}$
and $\nu_h$- turbulent viscosity in the horizontal direction.
Turbulent viscosity coefficients have been determined by Talon and Zahn (\cite{TZ97}), Maeder (\cite{M2003}), and Mathis et al. (\cite{MPZ04}).
Equations (3) and (4) are solved along with equations of stellar structure and evolution (Staritsin \cite{St99}, \cite{St05}, \cite{St07}, \cite{St14}).

   \begin{figure}
   \centering
   \includegraphics[width=\textwidth, angle=0]{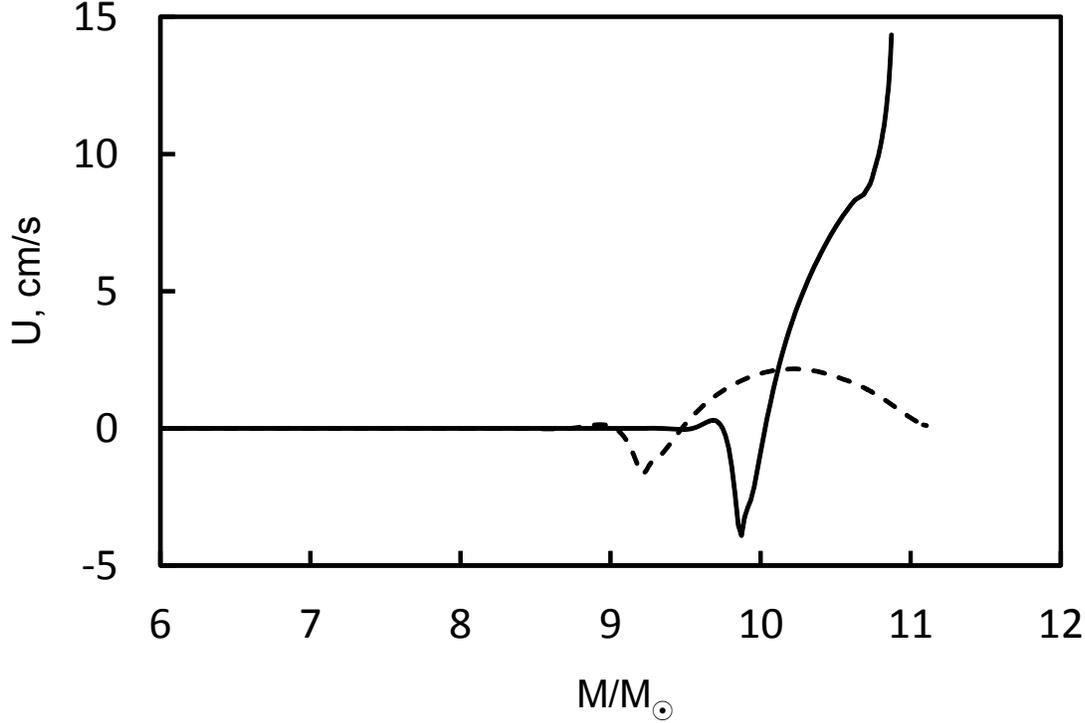}
   \caption{
   Velocity amplitude of meridional circulation before (solid line) and after (dashed line) the accretor receives critical rotation.
   }
   \label{Fig1}
   \end{figure}

The angular momentum transfer inside the accreting components of binary systems has been studied in two cases (Staritsin \cite{St19}, \cite{St21}). The case when a B star is in
the envelope of a red supergiant can be characterised by a high rate of increase in mass and angular momentum and a short accretion time (Staritsin \cite{St21}). Under the surface
of the B star, an outer circulation cell is created. The meridional circulation velocity in this cell is much greater than that of isolated stars. The bottom of the cell descends
into the star's interior and reaches the convective core before the red supergiant's envelope is ejected. Meridional circulation is the key mechanism for the transfer of angular
momentum inside a star. Shear turbulence is effective in a relatively narrow layer where the rate of angular rotation gradient has the greatest value. We also consider a particular
case of mass transfer in the Hertzsprung-gap of the HR diagram, when an intermediate convective zone is created in a more massive star above a shell source for the nuclear
burning of hydrogen. In this case, mass exchange on the nuclear time scale of helium burning occurs for the more massive star (Massevitch et al. \cite{MTY76}; Kraicheva et al. \cite{KTY77}; Staritsin \cite{St91}). As early as at the beginning of mass exchange, a single direction of circulation of matter in the accretor is established from the surface to the convective core. Meridional circulation is the main mechanism for the transfer of angular momentum into the interior of a star. The rate of angular rotation of the convective core and the lower part
of the radiative envelope increase and remain the same during the mass-transfer phase. In both considered cases, the meridional circulation plays a decisive role. In this paper,
we study the transfer properties of meridional circulation when the mass-transfer phase lasts on the thermal time scale. The transport properties of shear turbulence in the vertical direction are artificially suppressed.

\subsection{Features of the calculation of mass accretion}
\label{subsect:Features}
A main sequence star consists of a dense core and a rarefied envelope. Most of the stellar matter is concentrated in its core. The rarefied envelope accounts for an insignificant
fraction of the star's mass. When studying the evolution of stars, the structure of the core and rarefied envelope is calculated separately (Paczynski \cite{P1970}). In calculations
of the increase in the mass of the star due to accretion, we took a mass of the rarefied envelope equal to $0.035\,M_\odot$. The accreted mass is added to the core. An increase
in the core mass in the initial model should lead to the appearance of shells with low mass on its surface. Since an increase in stellar mass occurs on a thermal time scale,
the presence of low-mass shells requires small time steps. This complicates calculations of the accretor structure.

   \begin{figure}
   \centering
   \includegraphics[width=\textwidth, angle=0]{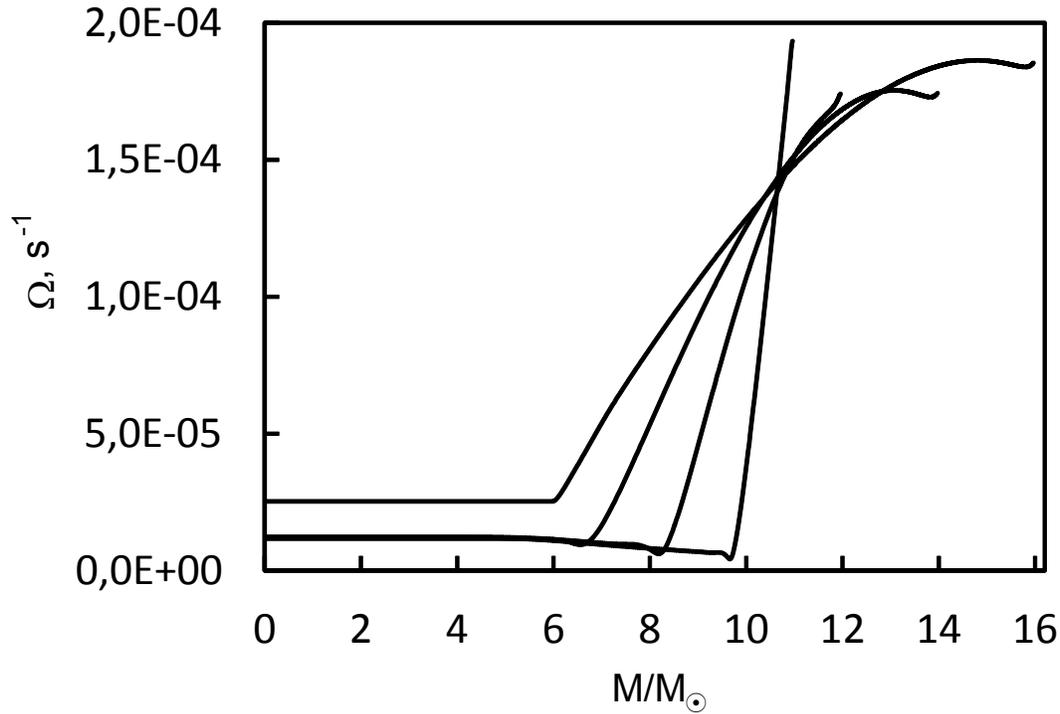}
   \caption{
   Distribution of the angular rate of rotation inside the accretor when its mass is 11 $M_\odot$, 12 $M_\odot$, 14 $M_\odot$, and 16 $M_\odot$.
   }
   \label{Fig2}
   \end{figure}

The size of the core is approximately 70\% of the star radius. The Keplerian rotation of the layer near the star surface gives the super-Keplerian rotation of this layer near the surface of the core, if the angular momentum of the layer is kept. The Keplerian rotation on the surface of the core corresponds to about 80\% of the Keplerian rotation on the star's surface. It is this stellar rotation that we have taken as the critical rotation. Thus, in the state of critical rotation, a layer of mass with Keplerian rotation is added to the core's surface. At the same time, the angular momentum entering the star is underestimated by about 20\%.

\section{Calculation data}
\label{sect:calculation}
By the beginning of mass exchange, the hydrogen content in the convective core of a star with a mass of $10.7\,M_\odot$ decreases to $\sim0.35$. This star is slowly rotating.
The non-spherical distribution of the average molecular weight suppresses meridional circulation in the layer of variable chemical composition above the convective core.
The meridional circulation carries the angular momentum outward in the chemically homogeneous envelope of the star. The circulation velocity is $\sim10^{-5}\,\mbox{cm}\,\mbox{s}^{-1}$.
Angular momentum transfer is not effective due to the star's slow rotation.

   \begin{figure}
   \centering
   \includegraphics[width=\textwidth, angle=0]{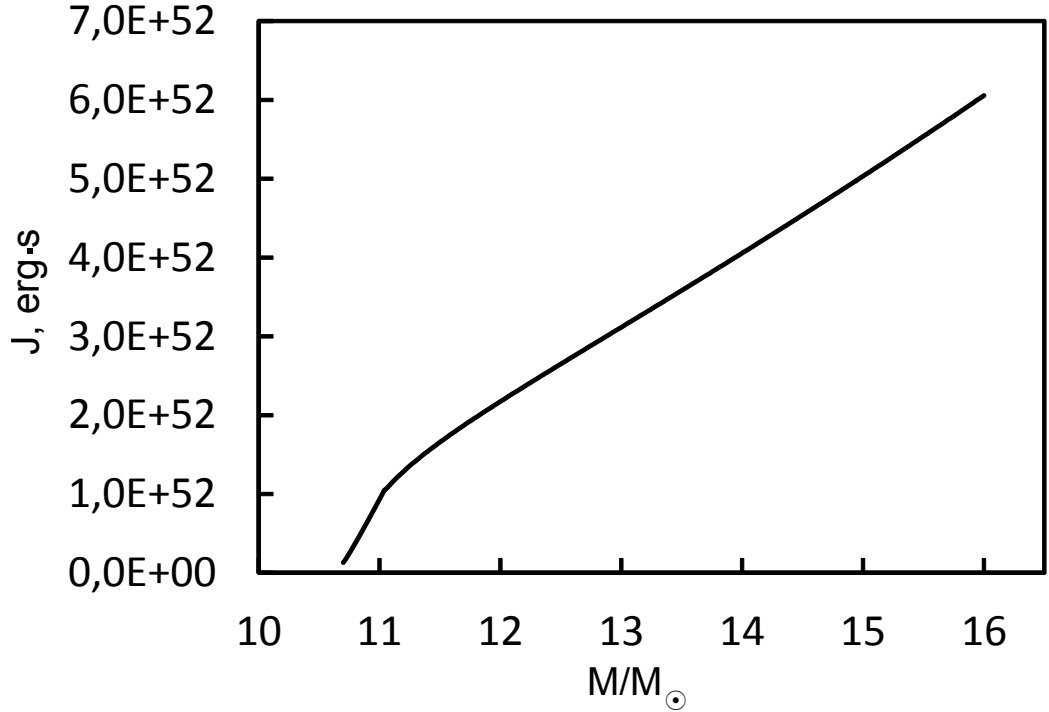}
   \caption{
   Angular momentum of an accretor as a function of its mass.
   }
   \label{Fig3}
   \end{figure}

At the very beginning of accretion, the angular momentum of the star increases for two reasons. First, the star acquires mass at the same rate of angular rotation as on the stellar surface. Secondly, angular momentum is introduced into the surface layer of the star by turbulence or waves with rate according to Equation~(1). It is this source of angular momentum that is
the main one at the beginning of accretion. Due to the input of the angular momentum, a circulation cell is created in the accretor's surface layer, in which the meridional circulation transfers the angular momentum into the accretor's interior. The meridional circulation velocity in this cell is  $\sim10\,\mbox{cm}\,\mbox{s}^{-1}$ (Fig.~1). This cell grows due to the addition of mass to the accretor and because the lower boundary of the cell drops down into the accretor. Below this cell, four more circulation cells with opposite directions of angular momentum transfer are created. The lower the cell is, the lower the maximum absolute value of the meridional circulation velocity in it.

When the mass of the accretor increases to $11\,M_\odot$, the rate of stellar rotation reaches a critical value. The subsequent increase in the angular momentum of the accretor occurs
only due to the addition of mass. The inflow of angular momentum into the surface layer of the star by means of waves or turbulence~(1) no longer occurs. The velocity of the meridional circulation in the outer cell decreases and becomes about $\sim1\,\mbox{cm}\,\mbox{s}^{-1}$ (Fig.~1).

During the remaining time of accretion, the star acquires another $5\,M_\odot$. Boundary condition according Equation~(2) determines the critical rotation of the stellar surface.
The angular momentum that enters the star along with the accreted matter is partially transferred by circulation to the deep inner layers. As a result, the rate of rotation of those
layers increases (Fig.~2). However, only the surface value of the rate of rotation coincides with the critical value. The rate of rotation of the inner layers remains much lower than
the local Keplerian value. When the bottom of the circulation cell, to which the angular momentum is transferred inside the star, descends to a layer with a variable chemical composition, the baroclinic component in the meridional circulation velocity becomes larger than the component due to the non-spherical distribution of the average molecular mass (Staritsin \cite{St05}, \cite{St07}). The meridional circulation extends to layers with variable chemical composition. When the star's mass increases to $15\,M_\odot$, the lower boundary of the cell drops to the convective core. After this, the angular momentum begins to flow into the convective core.

   \begin{figure}
   \centering
   \includegraphics[width=\textwidth, angle=0]{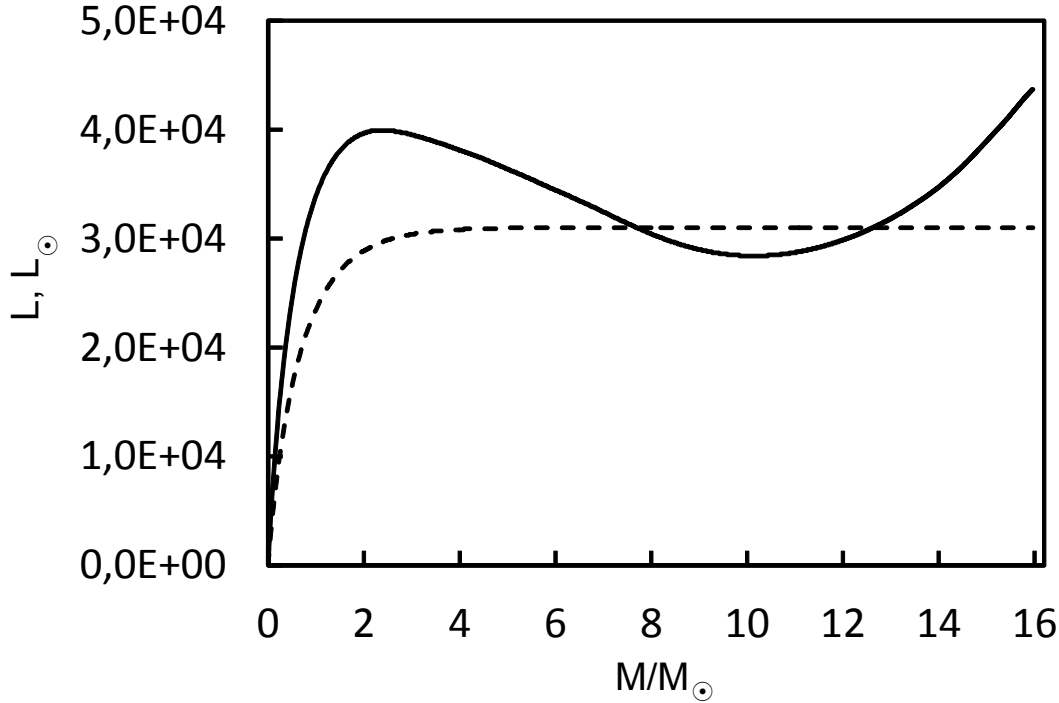}
   \caption{
   Luminosity distribution in the interior of the accretor immediately after the end of accretion (solid line)
   and after the expiration of the thermal time scale after the end of accretion (dashed line).
   }
   \label{Fig4}
   \end{figure}

The mass and angular momentum of the star continuously increase during accretion (Fig.~3). By the end of accretion, the star consists of two parts: a spun up inner part that made up
the star before the start of accretion and an outer part consisting of accreted matter. Most of the star's angular momentum is in the outer part. The stellar mass increases
to $16\,M_\odot$ and the angular momentum increases to $6.06\times10^{52}\,\mbox{erg}\cdot\mbox{s}$. The hydrogen content in the convective core increases to $\sim0.50$. The sequence
of stars of increasing mass in a state of critical rotation in the case of the accreting components of binary systems is a sequence of stars with increasing mass, increasing angular momentum, critical surface rotation, and subcritical rotation of the inner layers.

During accretion, the star goes out of thermal equilibrium. After the end of accretion, thermal equilibrium is restored on the thermal time scale (Fig.~4). At the very beginning of the restoration of thermal equilibrium, the radius of the star decreases rapidly. Calculations for the evolution of a star with a constant value of angular momentum lead to the supercritical rotation of the stellar surface. Therefore, the beginning of the restoration of thermal equilibrium is calculated with boundary condition (2). The star's angular momentum decreases to $5.97\times10^{52}\,\mbox{erg}\cdot\mbox{s}$. During the subsequent restoration of thermal equilibrium, the angular momentum of the star remains constant. Throughout this time,
the meridional circulation transfers the angular momentum from the accreted layers of the star to the inner ones. The angular velocity of rotation of the inner and outer layers of the star gradually equalises (Fig.~5). As a result, the rate of meridional circulation decreases.

   \begin{figure}
   \centering
   \includegraphics[width=\textwidth, angle=0]{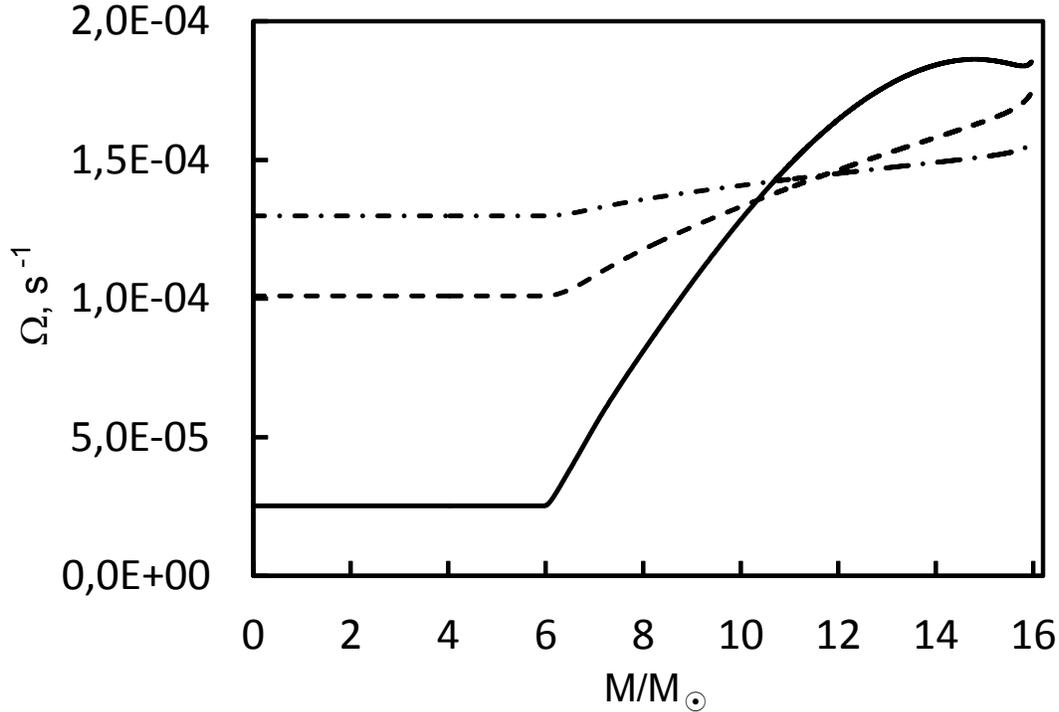}
   \caption{
   Distribution of the angular velocity of rotation in the interior of the accretor immediately after the end of accretion (solid line),
   in the middle (dashed line), and at the end of the thermal time scale (dash-dot line).
   }
   \label{Fig5}
   \end{figure}

As a result of mass transfer, the mass of the accretor increases by a factor of $\sim1.5$, and the angular momentum by $\sim50$ times. The angular momentum distribution in stellar interiors changes continuously during the mass-transfer phase and after it ends. The angular momentum in the inner part, which constituted the star before the start of accretion, continuously increases (Fig.~6). The increase in the angular momentum of this part after the end of accretion is approximately the same as during accretion. At the expiration of the thermal time scale after the end of accretion, this part of the star contains 30\% of the star's angular momentum. The transportation of the angular momentum that enters the star along with the accreted matter to the inner parts of the star requires a finite time. A strong increase in the velocity of the meridional circulation in response to the accretion of mass and angular momentum reduces the characteristic time of angular momentum transfer in the interior of the accretor compared to an isolated star, but still this time remains longer than the duration of the mass-transfer phase.

After the restoration of thermal equilibrium, the rotation of the close binary component still differs from the rotation of an isolated star with the same mass, angular momentum, and hydrogen content in the convective core. The rotation rate of the inner layers of the close binary component is slightly slower, and that of the outer layers is slightly higher, than that of an isolated star. The velocity of meridional circulation decreases to $\sim10^{-2}\,\mbox{cm}\,\mbox{s}^{-1}$ and remains higher than that of an isolated star. The circulation continues to transfer angular momentum from the outer layers of the component to the inner ones. The further evolution of the velocity field is also determined by the nuclear evolution of the component and tidal interaction.

\section{Discussion}
\label{sect:discussion}
The evolution of an isolated star with a mass of $16\,M_\odot$ and an angular momentum $J$ in the range $0.92\times10^{52}-3.69\times10^{52}\,\mbox{erg}\cdot\mbox{s}$ was studied on the main sequence (Staritsin \cite{St07}). The inner part of a star with a mass of $10.7\,M_\odot$ contains 47\% of the angular momentum of the star at $J=0.92\times10^{52}\,\mbox{erg}\cdot\mbox{s}$ and 34\% at $J=3.69\times10^{52}\,\mbox{erg}\cdot\mbox{s}$, when the hydrogen content in the convective core is the same as in the accretor. The rotation of the accretor after the expiration of the thermal time scale after the end of accretion is close to the rotation of an isolated star.

The evolution of an isolated star is accompanied by a transfer of angular momentum from the inner parts to the outside. The more intense the angular momentum transfer, the greater the angular momentum of the star. At $J\ge1.83\times10^{52}\,\mbox{erg}\cdot\mbox{s}$, the ratio of the surface rotation velocity at the equator to the Keplerian velocity $V_e/V_c$  slightly decreases at the beginning of evolution, then increases. At $J=3.69\times10^{52}\,\mbox{erg}\cdot\mbox{s}$, the $V_e/V_c$ ratio has values typical of Be-type stars during the second half of a star’s evolution on the main sequence. Extrapolation of the results of Staritsin (\cite{St07}) gives the minimum value of the ratio $V_e/V_c\sim0.95$
at $J=5.97\times10^{52}\,\mbox{erg}\cdot\mbox{s}$ on the main sequence.

   \begin{figure}
   \centering
   \includegraphics[width=\textwidth, angle=0]{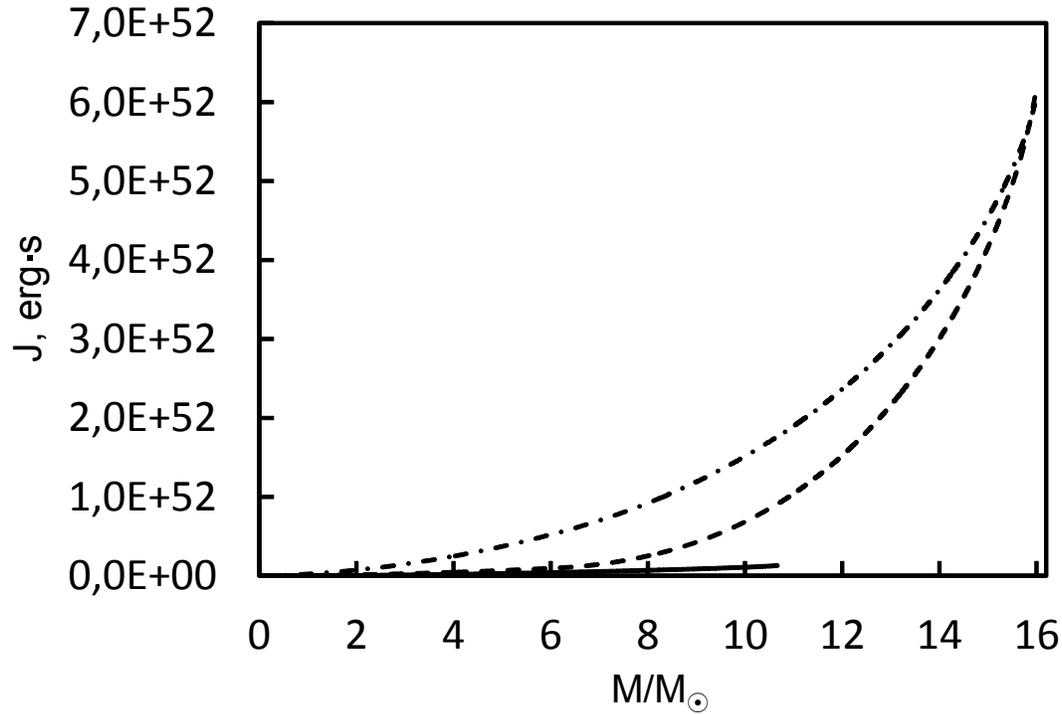}
   \caption{
   Distribution of angular momentum in the interior of the accretor before the start of accretion (solid line),
   after the end of accretion (dashed line) (note: the total angular momentum of the accretor is greatest at this point),
   and after the end of the thermal time scale after the end of accretion (dash-dot line).
   }
   \label{Fig6}
   \end{figure}

In the proposed calculation of mass exchange, the increase in the angular momentum of the accretor is underestimated by $\sim$20\%. However, the angular momentum of the accretor increased
to a value at which the ratio of the equatorial surface rotation velocity to the Keplerian velocity would not fall below $\sim0.95$ for isolated stars of the main sequence. Thus, a component of a close binary system can have the characteristics of a Be-type star immediately after the end of the mass-transfer phase in a close binary. However, unlike isolated stars, the angular momentum of the close binary component can decrease due to tidal interaction. The magnitude of this effect depends on the parameters of the close binary after the end of the mass-transfer phase. These parameters are determined by how much mass and angular momentum is lost from the close binary during the mass-transfer phase.

\section{Conclusions}
\label{sect:conclusion}
We have considered the spinning up of an initially less massive star in a close binary system in the case of typical mass transfer in the Hertzsprung-gap on the thermal time scale. The mass and angular momentum of this star increase due to the accretion of part of the matter and the angular momentum lost by the other star due to Roche lobe filling. The star reaches critical rotation after a $\sim$3\% increase in mass. Critical rotation can be characterised by the critical rate of surface rotation and the subcritical rotation of the inner layers. Further, due to accretion, the star moves along a sequence of stars of increasing mass with a critical rotation according to Bisnovatyi-Kogan (\cite{BK93}). During the mass-transfer phase, the mass of the star increases by a factor of $\sim1.5$, and the angular momentum increases by a factor of $\sim50$. Isolated stars with such an angular momentum can be characterised by rapid surface rotation: the equatorial velocity exceeds 95\% of the Keplerian velocity during the entire main sequence. Thus, mass transfer in a close binary could be the reason for the rapid rotation of Be stars.

We studied the transfer of angular momentum in the interior of a star by meridional circulation, which was taken into account according to the theory of Zahn (\cite{Zahn92}). The velocity
of meridional circulation increases by several orders of magnitude in response to the addition of mass and angular momentum to the star. The typical time of angular momentum transfer
by meridional circulation becomes much shorter than the evolution time of the star on the main sequence. However, only part of the acquired angular momentum enters the deep inner layers of the star during the mass-transfer phase. Another part of the acquired angular momentum is transferred inside the star after the end of the mass transfer for a finite time
on the thermal time scale.

\begin{acknowledgements}
This work was supported by the Ministry of Science and Education, FEUZ-2020-0030
\end{acknowledgements}

%

\label{lastpage}

\end{document}